\documentclass[aps, prd, showpacs, superscriptaddress, nofootinbib,
twocolumn]{revtex4}

\usepackage{amssymb, amsmath, bm, dcolumn, epsf, graphicx, latexsym, slashed}

\begin{document}

\title{Nonsingular bouncing cosmologies in light of BICEP2}

\author{Yi-Fu Cai}
\email{yifucai@physics.mcgill.ca}
\affiliation{Department of Physics, McGill University, Montr\'eal, QC, H3A
2T8, Canada}

\author{Jerome Quintin}
\email{jquintin@physics.mcgill.ca}
\affiliation{Department of Physics, McGill University, Montr\'eal, QC, H3A
2T8, Canada}

\author{Emmanuel N. Saridakis}
\email{Emmanuel\_Saridakis@baylor.edu}
\affiliation{Physics Division, National Technical University of Athens, 15780
Zografou Campus,  Athens, Greece}
\affiliation{Instituto de F\'{\i}sica, Pontificia Universidad de Cat\'olica
de Valpara\'{\i}so, Casilla 4950, Valpara\'{\i}so, Chile}

\author{Edward Wilson-Ewing}
\email{wilson-ewing@phys.lsu.edu}
\affiliation{Department of Physics and Astronomy, Louisiana State University,
Baton Rouge, 70803 USA}


\begin{abstract}
We confront various nonsingular bouncing cosmologies with the recently released BICEP2 data and investigate the observational constraints on their parameter space. In particular, within the context of the effective field approach, we analyze the constraints on the matter bounce curvaton scenario with a light scalar field, and the new matter bounce cosmology model in which the universe successively experiences a period of matter contraction and an ekpyrotic phase. Additionally, we consider three nonsingular bouncing cosmologies obtained in the framework of modified gravity theories, namely the Ho\v{r}ava-Lifshitz bounce model, the $f(T)$ bounce model, and loop quantum cosmology.
\end{abstract}

\maketitle

\section{Introduction}

Very recently, the BICEP2 collaboration announced the detection of primordial
B-mode polarization in the cosmic microwave background (CMB), claiming an
indirect observation of gravitational waves. This result, if confirmed by
other collaborations and future observations, will be of major significance
for cosmology and theoretical physics in general. In particular, the BICEP2
team found a tensor-to-scalar ratio \cite{Ade:2014xna}
\begin{equation}
\label{rBICEP2}
r=0.20_{-0.05}^{+0.07},
\end{equation}
at the $1\sigma$ confidence level for the $\Lambda$CDM scenario. Although
there remains the possibility that the observed B-mode polarization could be
partially caused by other sources \cite{Lizarraga:2014eaa, Moss:2014cra,
Bonvin:2014xia}, it is indeed highly probable that the observed B-mode
polarization in the CMB is due at least in part to gravitational waves,
remnants of the primordial universe.

The relic gravitational waves generated in the very early universe is a
generic prediction in modern cosmology \cite{Grishchuk:1974ny,
Starobinsky:1979ty}. Inflation is one of several cosmological paradigms that
predicts a roughly scale-invariant spectrum of primordial gravitational waves
\cite{Starobinsky:1979ty, Rubakov:1982df, Starobinsky:1985ww}. The same
prediction was also made by string gas cosmology \cite{Brandenberger:1988aj,
Brandenberger:2006xi, Brandenberger:2014faa, Biswas:2014kva} and the matter bounce scenario
\cite{Wands:1998yp, Finelli:2001sr, Cai:2008qw}. (Note that the specific
predictions of $r$ and the tilt of the tensor power spectrum can be used in
order to differentiate between these cosmologies.) So far, a lot of the
theoretical analyses of the observational data have been in the context of
inflation (see, for instance, \cite{Kehagias:2014wza, Ma:2014vua,
Harigaya:2014qza, Gong:2014qga, Miranda:2014wga,
Hertzberg:2014aha,Lyth:2014yya, Xia:2014tda, Hazra:2014jka,
Cai:2014bda, Hossain:2014coa, Hu:2014aua, Zhao:2014rna, Zhang:2014dxk,
DiBari:2014oja, Li:2014cka, Chung:2014woa, Cai:2014hja,Hossain:2014ova}).

In this present work, we are interested in exploring the consequences of the
BICEP2 results in the framework of bouncing cosmological models. In
particular, we desire to study the production of primordial gravitational
waves in various bouncing scenarios, in both the settings of effective field
theory and modified gravity. First, we show that the tensor-to-scalar ratio
parameter obtained in a large class of nonsingular bouncing models is
predicted to be quite large compared with the observation. Second, in some
explicit models this value can be suppressed due to the nontrivial physics of
the bouncing phase, namely, the matter bounce curvaton \cite{Cai:2011zx} and
the new matter bounce cosmology \cite{Cai:2012va, Cai:2013kja, Cai:2014bea}.
Additionally, for bounce models where the fluid that dominates the
contracting phase has a small sound velocity, primordial gravitational waves
can be generated with very low amplitudes \cite{Bessada:2012kw}. We show
that the current Planck and BICEP2 data constrain 
the energy scale at which the bounce occurs as well as the slope of the
Hubble rate during the bouncing phase in these specific models.

The paper is organized as follows. In Sec.\ \ref{s.matt-bounce}, we focus
on matter bounce cosmologies from the effective field theory perspective. In
particular, we explore the matter bounce curvaton scenario \cite{Cai:2011zx}
and the new matter bounce cosmology \cite{Cai:2013kja}. In
Sec.\ \ref{s.mod-grav}, we explore another avenue for obtaining nonsingular
bouncing cosmologies, that is modifying gravity. We comment on the status of
the matter bounce scenario in Ho\v{r}ava-Lifshitz gravity
\cite{Brandenberger:2009yt}, in $f(T)$ gravity \cite{Cai:2011tc}, and in loop
quantum cosmology \cite{Cai:2014zga}. We conclude with a discussion in
Sec.\ \ref{s.concl}.

\section{Matter Bounce Cosmology}
\label{s.matt-bounce}

As an alternative to inflation, the matter bounce cosmology can also give
rise to scale-invariant power spectra for primordial density fluctuations and
tensor perturbations \cite{Wands:1998yp,Finelli:2001sr}. In the context of
the original matter bounce cosmology, both the scalar and tensor modes of
primordial perturbations grow at the same rate in the contracting phase
before the bounce. As a result, this naturally leads to a large amplitude of
primordial tensor fluctuations \cite{Cai:2008qw}, greater than the
observational upper bound. However, an important issue that has to be
additionally incorporated in these calculations is how the perturbations pass
through the bouncing phase, which can drastically decrease the
tensor-to-scalar ratio. One example is the matter bounce in loop quantum
cosmology, where the tensor-to-scalar ratio is suppressed by quantum gravity
effects during the bounce \cite{WilsonEwing:2012pu}.  Also, for some
parameter choices in the new matter bounce model, $r$ can be suppressed by a 
small sound speed of the matter fluid \cite{Cai:2013kja}.

In this section, we focus on two particularly interesting matter bounce
cosmological models in the effective field theory setting. First, we consider
the matter bounce curvaton model, in which the primordial curvature
perturbation can be generated from the conversion of entropy fluctuations
seeded by a second scalar field \cite{Cai:2011zx}. Secondly, we investigate
the new matter bounce cosmology, where the primordial curvature perturbations
can achieve a gravitational amplification  during the bounce
\cite{Cai:2012va}.

\subsection{The matter bounce curvaton model}

The matter bounce curvaton model was originally studied to examine whether
the bouncing solution of the universe is stable against possible entropy
fluctuations \cite{Cai:2011zx} and particle creation \cite{Cai:2011ci}. In a
toy model studied in \cite{Cai:2011zx}, a massless entropy field $\chi$ is
introduced such that it couples to the bounce field $\phi$ via the
interaction term $g^2\phi^2\chi^2$. The entropy field evolves as a tracking
solution in the matter contracting phase and its field fluctuations are
nearly scale-invariant provided the coupling parameter $g^2$ is sufficiently
small. The amplitude of this mode is comparable to the tensor modes and
scales as the absolute value of the Hubble parameter $H$ before the bounce.
Afterwards, the universe enters the bouncing phase and the kinetic term of
the entropy field varies rapidly in the vicinity of the bounce. As in the
perturbation equation of motion this term effectively contributes a
tachyonic-like mass, a controlled amplification of the entropy 
modes can be achieved in the bouncing phase. Since this term does not appear
in the equation of motion for tensor perturbations, the amplitude of
primordial gravitational wave is conserved through this phase. After the
bounce, the entropy modes will be transferred into curvature perturbations,
and this increases the amplitude of the power-spectrum of the primordial
density fluctuations. An important consequence of this mechanism is its
suppression of the tensor-to-scalar ratio.

In this subsection, we briefly review the analysis of \cite{Cai:2011zx}, in
light of the BICEP2 results. In the simplest version of the matter bounce
curvaton mechanism, there are only three significant model parameters, namely
the coupling parameter $g^2$, the slope parameter of the bouncing phase
$\Upsilon$ (which is defined by $H\equiv \Upsilon t$ around the bounce), and
the maximal value of the Hubble parameter $H_B$. The value of $H_B$ is
associated with the mass of the bounce field $m$ through the following
relation
\begin{eqnarray}\label{H_m}
 H_B \simeq \frac{4 m}{3\pi}~.
\end{eqnarray}

The propagation of primordial gravitational waves depends only on the
evolution of the scale factor, and it is possible to calculate the power
spectrum for primordial gravitational waves in this scenario%
\footnote{$M_p\equiv 1/ \sqrt{8\pi G}$ is the reduced Planck mass.}
\begin{eqnarray}
 P_T = \frac{2H_\mathrm{m}^2}{9\pi^2 M_p^2} ~,
\end{eqnarray}
from which we see that the amplitude is determined solely by the maximal
Hubble scale $H_\mathrm{m}$. However, the amplitude of the entropy fluctuations is
increased during the bounce. Since tensor perturbations do not couple to
scalar perturbations, the entropy perturbations do not affect the power
spectrum of gravitational waves, whereas the entropy modes are amplified and
act as a source for curvature perturbations. This asymmetry leads to a
smaller tensor-to-scalar ratio of
\begin{eqnarray}
 r \simeq \frac{35}{{\cal F}^2}~,
\end{eqnarray}
where the amplification factor is given by
\begin{eqnarray}
 {\cal F} \simeq
e^{\sqrt{y(2+y)}+\frac{3}{\sqrt{2}}\sinh^{-1}(\frac{2\sqrt{y}}{3})} ~,~ {\rm
with}~ y\equiv \frac{m^2}{\Upsilon}~,
\end{eqnarray}
and $\Upsilon$ is the slope parameter of the bouncing phase as defined before
Eq.\ \eqref{H_m}. Since the exponent in the above equation is approximately
linear in $y$ in the regime of interest, we see that the tensor-to-scalar
ratio can be greatly suppressed for large values of $y$, that is for large
$m$ or small $\Upsilon$. Also, we see that $r$ will reach a maximal value in
the massless limit or in the limit where the bounce is instantaneous (i.e.,
$\Upsilon\rightarrow\infty$), in which case entropy perturbations are not
enhanced.

\begin{figure}
\includegraphics[scale=0.6]{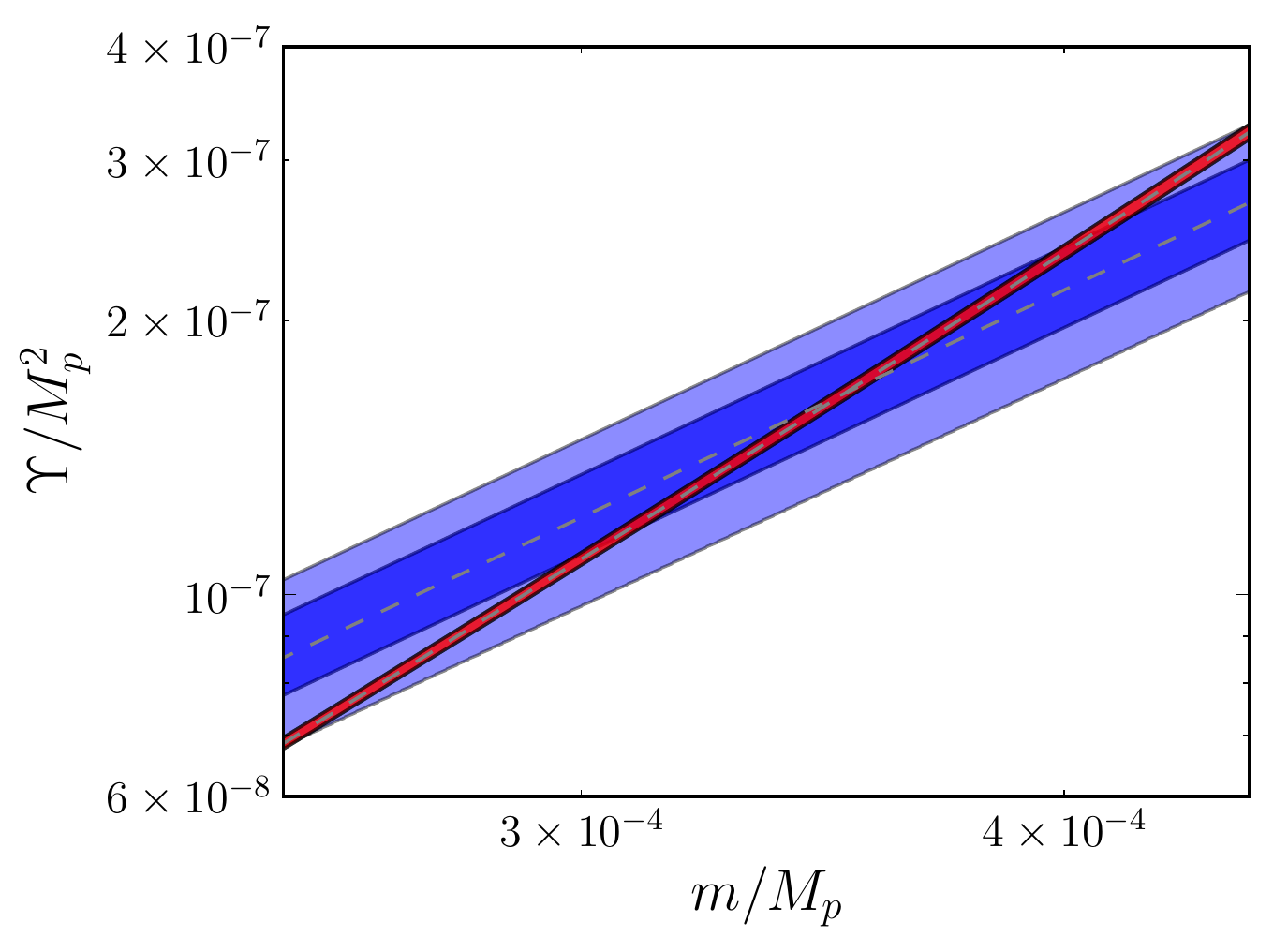}
\caption{Constraints on the mass parameter $m$ and the slope parameter $\Upsilon$ of the bounce phase from the measurements of Planck and BICEP2 in the matter bounce curvaton scenario. The blue bands show the 1$\sigma$ and 2$\sigma$ confidence intervals of the tensor-to-scalar ratio and the red bands show the confidence intervals of the amplitude of the power spectrum of curvature perturbations.}
\label{fig1}
\end{figure}

We recall that, according to the latest observation of the CMB (Planck+WP),
the amplitude of the power spectrum of primordial curvature perturbations is
constrained to be \cite{Ade:2013zuv}
\begin{eqnarray}
 \ln(10^{10}A_s) = 3.089^{+0.024}_{-0.027} ~(1\sigma ~{\rm CL})~,
\end{eqnarray}
at the pivot scale $k=0.002\,{\rm Mpc}^{-1}$. Moreover, the recently released
BICEP2 data indicate that \cite{Ade:2014xna}
\begin{eqnarray}
 r = 0.20^{+0.07}_{-0.05} ~(1\sigma ~{\rm CL})~.
\end{eqnarray}
By making use of the above data, we performed a numerical estimate and
derived the constraint on the model parameters $\Upsilon$ and $m$ shown in
Fig.\ \ref{fig1}. From the result, we find that the mass scale $m$ and the
slope parameter $\Upsilon$ appearing in the matter bounce curvaton model have
to be in the following ranges
\begin{align}
 2.5\times 10^{-4} \lesssim m/M_p \lesssim 4.5\times 10^{-4} ~, \\
 7.0\times 10^{-8} \lesssim \Upsilon/M_p^2 \lesssim 3.5\times 10^{-7} ~,
\end{align}
respectively. The resulting constraints suggest that if the universe has
experienced a nonsingular matter bounce curvaton, then the energy scale of
the bounce should be of the order of the GUT scale with a smooth and slow
bouncing process.

\subsection{New matter bounce cosmology}
\label{ss.new-mb}

In the new matter bounce scenario, as first developed in \cite{Cai:2012va},
the universe starts with a matter-dominated period of contraction and evolves
into an ekpyrotic phase before the bounce. This scenario combines the
advantages of the matter bounce cosmology, which gives rise to
scale-invariant primordial power spectra, and the ekpyrotic universe, which
strongly dilutes primordial anisotropies \cite{Cai:2013vm}. The model can be
implemented by introducing two scalar fields, as analyzed in the context of
effective field theory \cite{Cai:2013kja}.

In the effective model of the two field matter bounce \cite{Cai:2013kja}, one
scalar field is introduced to drive the matter-dominated contracting phase
and the other is responsible for the ekpyrotic phase of contraction and the
nonsingular bounce. Therefore, similar to the matter bounce curvaton
scenario, there also exist curvature perturbations and entropy fluctuations
during the matter-dominated contracting phase. However, the main difference
between these two models is that,
in the present case, the entropy modes have already been converted into
curvature perturbation when the universe enters the ekpyrotic phase before
the bounce, while in the matter bounce curvaton mechanism, this process
occurs after the bounce.

In this model, when the universe evolves into the bouncing phase, the kinetic
term of the scalar field that triggers the bounce could vary rapidly which is
similar to the analysis of the matter bounce curvaton mechanism. This process
can also effectively lead to a tachyonic-like mass for curvature
perturbations, and therefore, the corresponding amplitude can be amplified.
For the same reason as the matter bounce curvaton mechanism, this effect only
works on the scalar sector. Correspondingly, the tensor-to-scalar ratio is
suppressed when primordial perturbations pass through the bouncing phase in
the new matter bounce cosmology. We would like to point out that this effect
is model dependent, namely, it could be secondary if the kinetic term of the
background scalar evolves very smoothly compared to the bounce phase
\cite{Osipov:2013ssa, Koehn:2013upa}.

Following \cite{Cai:2013kja}, one can write the expression of the power
spectrum for primordial tensor fluctuations as
\begin{eqnarray}
 P_T \simeq \frac{{\cal F}_{\psi}^2\gamma_\psi^2
H_\mathrm{E}^2}{16\pi^2(2q-3)^2M_p^2}~,
 \label{PT}
\end{eqnarray}
with
\begin{align}
 \gamma_\psi &\simeq \frac{1}{2(1-3q)}~, \nonumber\\
 {\cal F}_{\psi} &\simeq \exp \left[ 2\sqrt{\Upsilon}t_{B+}
+\frac{2}{3}\Upsilon^{3/2}t_{B+}^3 \right]~,
\end{align}
up to leading order. In the above expression, $H_\mathrm{E}$ is the value of the
Hubble rate at the beginning of the ekpyrotic phase and $q$ is an ekpyrotic
parameter which is much less than unity. Note that we have assumed that the
bouncing phase is nearly symmetric around the bounce point with the values of
the scale factor before and after the bounce being comparable. We denote the
time at the end of the bounce phase by $t_{B+}$.

At leading order, the power spectrum for curvature fluctuations is given by
\begin{equation}
 P_\zeta\simeq\frac{\mathcal{F}_{\zeta}^2H_\mathrm{E}^2a_\mathrm{E}^2}{8\pi^2M_p^4}\gamma_{
\zeta}^2m^2\left|U_{\zeta}\right|^2~,
 \label{Pz}
\end{equation}
with $\gamma_\zeta\simeq\gamma_\psi$ and
\begin{align}
  U_\zeta &= -(25+49q)i\frac{H_\mathrm{E}}{24m}-\frac{27q}{24}~,\nonumber \\
  {\cal F}_{\zeta} &\simeq e^{2\sqrt{2+\Upsilon
T^2}\left(\frac{t_{B+}}{T}\right) +\frac{2(2+3\Upsilon
T^2+\Upsilon^2T^4)}{3\sqrt{2+\Upsilon T^2}}
\left(\frac{t_{B+}^3}{T^3}\right)}~.
\end{align}
Similarly to $H_\mathrm{E}$, we introduced $a_\mathrm{E}$, which is the value of the scale
factor at the beginning of the ekpyrotic phase (in the pre-bounce branch of
the universe). Also, we introduced the mass $m$ of the scalar field
responsible for the phase of matter contraction. We also note the presence of
the variable $T$, which comes into play in the evolution of the bounce field
(see \cite{Cai:2013kja} for more details). From eqs.\ (\ref{PT}) and
(\ref{Pz}), the tensor-to-scalar ratio in this model then takes the form of
\begin{equation}
 r \equiv\frac{P_T}{P_\zeta} \simeq \frac{{\cal F}_{\psi}^2M_p^2}{2(2q-3)^2
\mathcal{F}_{\zeta}^2a_\mathrm{E}^2m^2 \left|U_{\zeta}^{(k)}  \right|^2}~.
\end{equation}

Similarly to the previous subsection, we perform a numerical estimate to
investigate the consequences of the observational constraints on the
parameter space of the new matter bounce scenario. Note that, although the
model under consideration
involves a series of parameters, there are three main parameters that are
most sensitive to observational constraints, i.e., the slope parameter
$\Upsilon$, the Hubble rate at the beginning of the ekpyrotic phase $H_\mathrm{E}$,
and the dimensionless duration parameter $t_{B+}/T$ of the bouncing phase.

\begin{figure}
\includegraphics[scale=0.6]{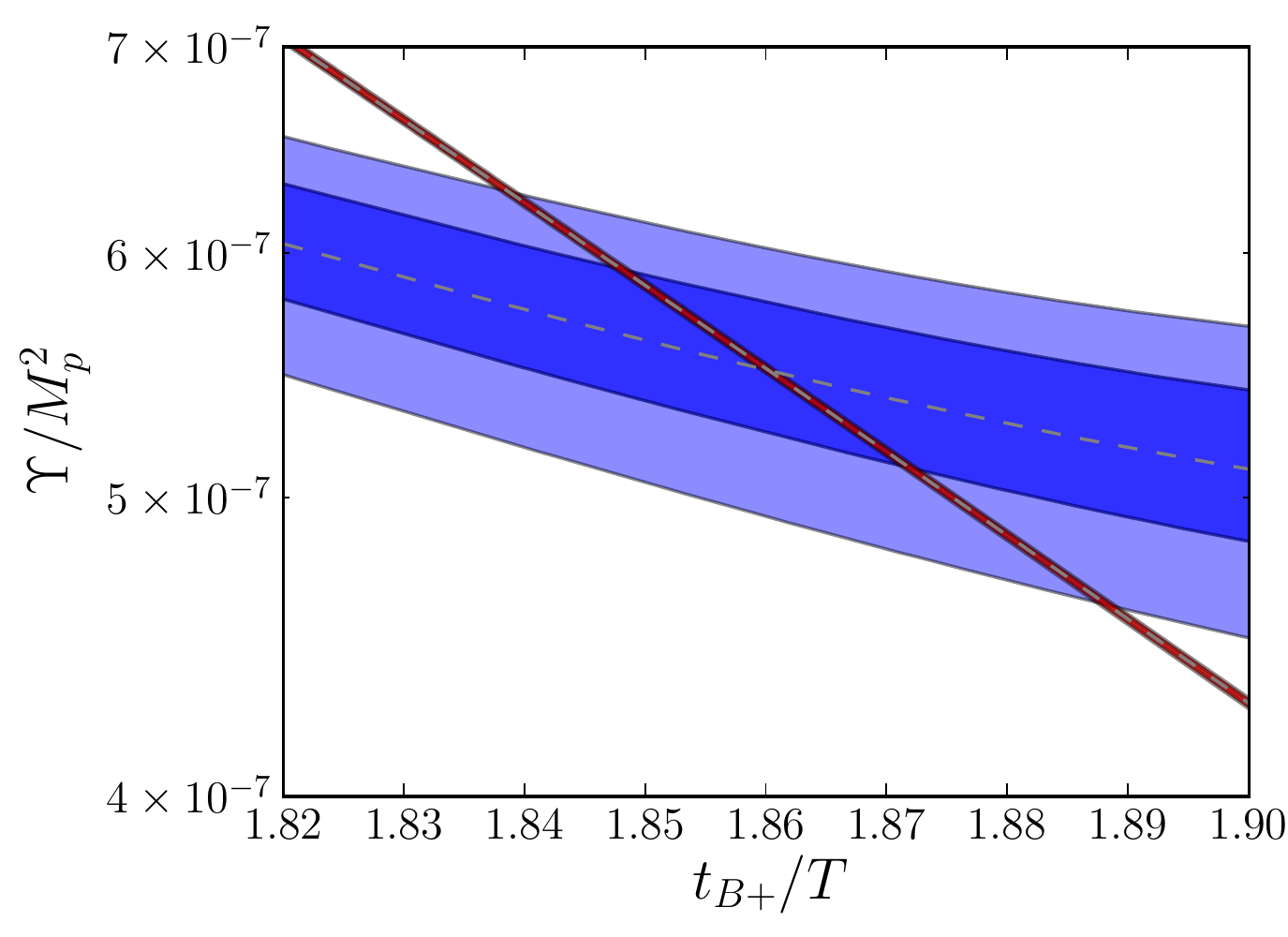}
\caption{Constraints on the dimensionless duration parameter $t_{B+}/T$ and the slope parameter $\Upsilon$ of the bounce phase from the measurements of Planck and BICEP2 in the new matter bounce cosmology. The value of the Hubble parameter at the beginning of the ekpyrotic phase is fixed to be $H_\mathrm{E}/M_p=10^{-7}$. As in Fig.\ \ref{fig1}, the blue and red bands show the confidence intervals of the tensor-to-scalar ratio and of the amplitude of the power spectrum of curvature perturbations, respectively.}
\label{Fig:nmb2}
\end{figure}

We first look at the correlation between $\Upsilon$ and $t_{B+}/T$, with the
numerical result shown in Fig.\ \ref{Fig:nmb2}. One can read that the slope
parameter $\Upsilon$ and the dimensionless duration parameter $t_{B+}/T$ are
slightly negatively correlated. This implies that one expects either a slow
bounce with a long duration or a fast bounce with a short duration. However,
it is easy to find that the constraint on the dimensionless duration
parameter $t_{B+}/T$ is very tight with a value slightly less than $2$.
Therefore, it is important to examine whether the model predictions
accommodate with observations by fixing the parameter $t_{B+}/T$.

\begin{figure}
\includegraphics[scale=0.6]{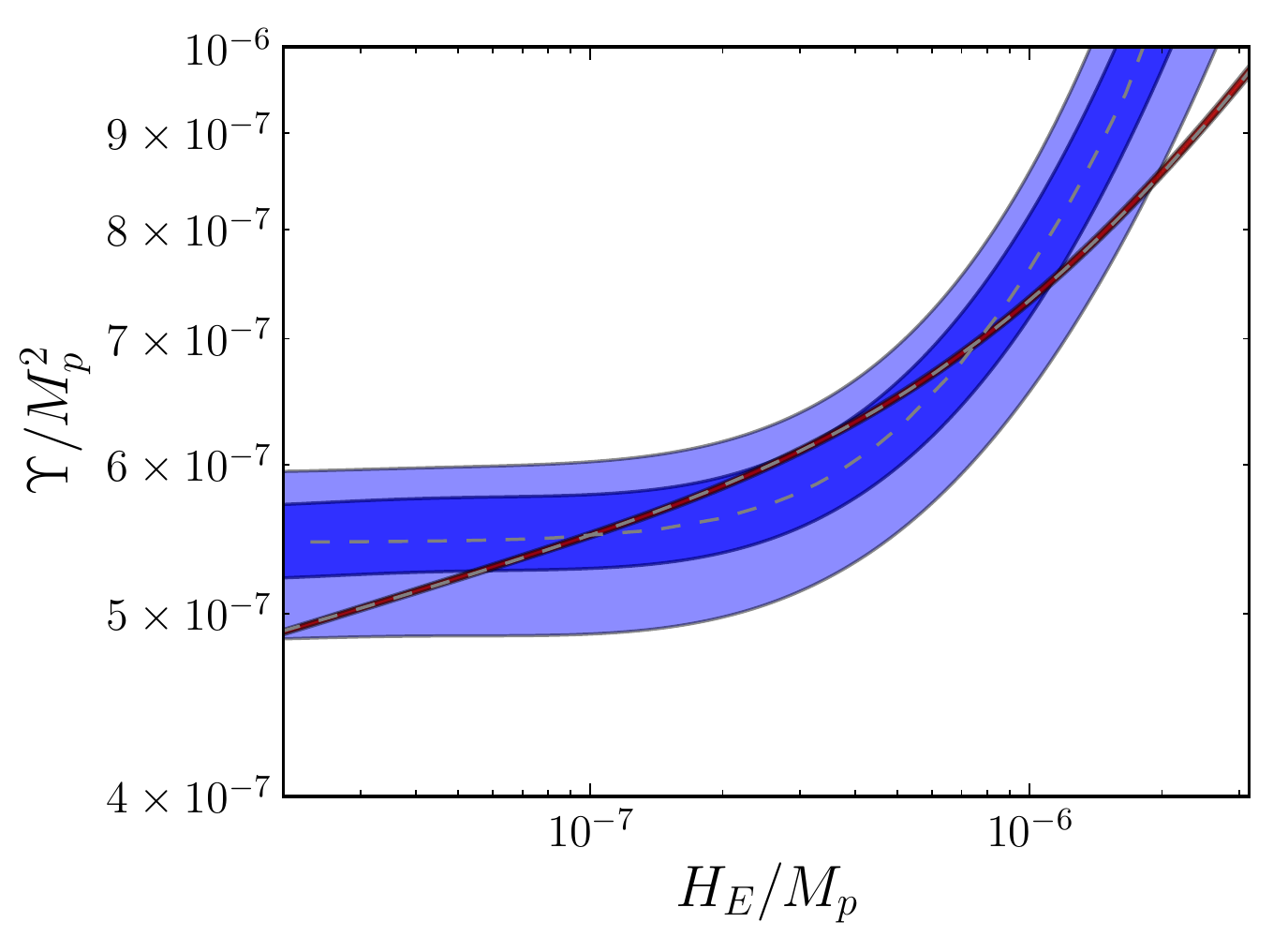}
\caption{Constraints on the Hubble parameter at the beginning of the ekpyrotic phase $H_\mathrm{E}$ and the slope parameter $\Upsilon$ of the bounce phase from the measurements of Planck and BICEP2 in the new matter bounce cosmology. The dimensionless bounce time duration is fixed to be $t_{B+}/T=1.86$. As in Figs.\ \ref{fig1} and \ref{Fig:nmb2}, the blue and red bands show the
confidence intervals of $r$ and of the amplitude of $P_\zeta$, respectively.}
\label{Fig:nmb3}
\end{figure}

Then, we analyze the correlation between $\Upsilon$ and $H_\mathrm{E}$ after setting $t_{B+}/T=1.86$. The allowed parameter space is depicted by the intersection of the blue and red bands shown in Fig. \ref{Fig:nmb3}. From the result, we find that the Hubble scale $H_\mathrm{E}$ and the slope parameter $\Upsilon$ introduced in the new matter bounce cosmology are constrained to be in the following ranges
\begin{align}
 1.9 \times 10^{-8} &\lesssim H_\mathrm{E}/M_p \lesssim 1.9 \times 10^{-6} ~, \\
 4.9 \times 10^{-7} &\lesssim \Upsilon/M_p^2 \lesssim 8.5 \times 10^{-7} ~,
\end{align}
respectively. One can easily see that the constraints on the slope parameter
$\Upsilon$ in the new matter bounce cosmology and in the matter bounce
curvaton scenario are in the same ballpark, i.e., $\Upsilon\sim
\mathcal{O}(10^{-7})$. For the new matter bounce cosmology, if we assume that
the bounce occurs at the GUT scale, then the duration of the bouncing phase
is roughly $\mathcal{O}(10^4)$ Planck times. We also note that the amplitude
of the Hubble scale $H_\mathrm{E}$ in the new matter bounce cosmology is much lower
than the GUT scale. This allows for a long enough ekpyrotic contracting phase
that can suppress the unwanted primordial anisotropies as addressed in
\cite{Cai:2013vm}.

In summary, from the analysis of the matter bounce curvaton and the new
matter bounce cosmology scenarios, we can conclude that, in general, a
nonsingular bouncing cosmology has to experience the bouncing phase smoothly
for it to agree with latest observational data. In other words, the Hubble
parameter cannot grow too fast during the bounce phase since the constraints
that we find favor a small value of $\Upsilon$. Depending on the detailed
bounce mechanism, the observed amplitude of the spectra of the CMB
fluctuations may be determined by the bounce scale or the value of the Hubble
parameter at the moment when primordial perturbations were frozen at
super-Hubble scales. For the matter bounce curvaton, the mass scale of the
bounce is of the order of $\mathcal{O}(10^{-4})\,M_p$ which is close to or
slightly lower than the GUT scale ($\mathcal{O}(10^{16})\,{\rm GeV}$). On the
other hand, for the new matter bounce cosmology, due to the introduction of
the ekpyrotic phase, it is the Hubble parameter $H_\mathrm{E}$ at 
the onset of the ekpyrotic phase that determines the amplitude of the
primordial spectrum of the perturbations, and it must be much lower than the
GUT scale in order to agree with observations. These interesting results
encourage further study of bouncing cosmologies following the effective field
approach.

\section{Implications for modified-gravity bouncing cosmology}
\label{s.mod-grav}

In the previous section, we performed numerical computations to constrain two
representative bounce cosmologies that are described by the effective field
approach. It is interesting to extend the analysis to bouncing cosmology
models where the bounce occurs due to modified-gravity theories.
In the following, we shall focus on three specific models.
The first one is to obtain the matter bounce
solution in the framework of a non-relativistic modification to Einstein
gravity,
namely the Ho\v{r}ava-Lifshitz bounce model \cite{Brandenberger:2009yt,
Cai:2009in}. The second is to realize the nonsingular bounce by virtue of
torsion gravity, i.e., the $f(T)$ bounce model \cite{Cai:2011tc}. And the
third is a study of the new matter bounce cosmology in the setting of loop
quantum cosmology \cite{Cai:2014zga}.

\subsection{Matter bounce in Ho\v{r}ava-Lifshitz gravity}

Ho\v{r}ava-Lifshitz gravity is argued to be a potentially UV complete
theory for quantizing the graviton, and it has important implications
in the physics of the very early universe. In particular, a nonsingular
bouncing
solution can be achieved in this theory when a non-vanishing spatial
curvature term is taken into account \cite{Calcagni:2009ar, Kiritsis:2009sh}.
In this case, the higher order spatial derivative terms of the gravity
Lagrangian can effectively contribute a stiff fluid with negative energy,
which can trigger the nonsingular bounce as well as
suppressing unwanted primordial anisotropies. Thus, the bouncing
solutions obtained in this picture are marginally stable against the BKL
instability.

As was shown in \cite{Brandenberger:2009yt}, if the contracting phase is
dominated by a pressure-less matter fluid, Ho\v{r}ava-Lifshitz gravity
can provide a realization of the matter bounce scenario. Moreover, for
primordial perturbations in the infrared limit, the corresponding power
spectrum for both the scalar and tensor modes are almost scale-invariant
\cite{Cai:2009hc}. However, the paradigm derived in this framework belongs
to the traditional matter bounce cosmology, and so the amplitude of
the tensor power spectrum is of the same order as the scalar power spectrum.
In this regard, the corresponding tensor-to-scalar ratio is too large to
explain the latest cosmological observations. To address this issue, one
needs to enhance the amplitude of the curvature perturbations generated
in the contracting phase in the infrared limit, for example by applying
the matter bounce curvaton mechanism.

\subsection{The $f(T)$ matter bounce cosmology}

We briefly describe the realization of the matter bounce in the $f(T)$
gravitational modifications of general relativity.
The $f(T)$ gravity theory is a generalization of
the formalism of the teleparallel equivalent of general relativity
\cite{Linder:2010py, Chen:2010va, Dent:2011zz}, in which one uses the
curvature-less Weitzenb{\"{o}}ck connection instead of the torsion-less
Levi-Civita one, and thus all of the gravitational information is included in
the torsion tensor.

We use the vierbeins ${\mathbf{e}_A(x^\mu)}$ (Greek indices run over the
coordinate space-time and capital Latin indices
run over the tangent space-time) as the dynamical field, related to the
metric through $g_{\mu\nu}(x)=\eta_{AB}\, e^A_\mu (x)\, e^B_\nu (x)$, with
$\eta_{AB} = \rm{diag} (1,-1,-1,-1)$. Thus, the torsion tensor is
${T}^\lambda_{\:\mu\nu} = e^\lambda_A \:(\partial_\mu e^A_\nu-\partial_\nu
e^A_\mu)$, and the torsion scalar is given by
\begin{eqnarray}
 T &=&
\frac{1}{4}T^{\mu\nu\lambda}T_{\mu\nu\lambda}+\frac{1}{2}T^{\mu\nu\lambda}
 T_{\lambda\nu\mu}-T_{\nu}^{\,\,\,\nu\mu}T^{\lambda}_{\,\,\,\lambda\mu} ~.
\label{Tquad}
\end{eqnarray}
Thus, inspired by the $f(R)$ modifications of Einstein-Hilbert action, one
can construct the $f(T)$ modified gravity by taking the gravity action to be
an arbitrary function of the torsion scalar through
$S = \int d^4x \, e [T+f(T) ] /16\pi G$.
One interesting feature of the $f(T)$ gravity is that the null energy
condition can be effectively violated, and thus nonsingular bouncing
solutions are possible. In particular, it has been shown that
the matter bounce cosmology can be achieved by reconstructing the form of
$f(T)$ under specific parameterizations of the scale factor
\cite{Cai:2011tc}.

In this model the power spectrum of primordial curvature perturbations is
also scale-invariant if the contracting branch is matter-dominated.
Its form is given by
\begin{eqnarray}
 P_{\zeta} = \frac{H_\mathrm{m}^2}{48\pi^2M_p^2}~,
\end{eqnarray}
where $H_\mathrm{m}$ is the maximal value of the Hubble parameter throughout the whole
evolution \cite{Cai:2011tc} (and thus its definition is the same as that
introduced in the matter bounce curvaton scenario).

Now we investigate the evolution of primordial tensor fluctuations in the
$f(T)$ matter bounce. The perturbation equation for the tensor modes can be
expressed as \cite{Chen:2010va}
\begin{equation}
\bigg(\ddot{h}_{ij}+3H\dot{h}_{ij}-\frac{\nabla
^{2}}{a^{2}}h_{ij}\bigg)-%
\frac{12H\dot{H}f_{,TT}}{1+f_{,T}}\dot{h}_{ij}=0~,  \label{h_eom}
\end{equation}%
where the tensor modes are transverse and traceless. It is interesting to
note that the last term appearing in Eq.\ \eqref{h_eom} plays a role of an
effective ``mass" for the tensor modes which may affect their amplitudes
along the cosmic evolution. However, as it was pointed out in
\cite{Cai:2011tc},
the effect brought by $f_{,TT}$ is negligible since $f(T)$ is approximately a
linear function of $T$ in the matter contracting phase. Hence, for primordial
tensor fluctuations at large length scales, although the power spectrum is
also scale-invariant, the amplitude takes of the form of
\begin{eqnarray}
 P_T = \frac{H_\mathrm{m}^2}{2\pi^2 M_p^2} ~.
\end{eqnarray}
Thus, this result is already ruled out by the present observations unless one
introduces some mechanism to magnify the amplitude of scalar-type metric
perturbations. This issue can be resolved by the matter bounce curvaton
mechanism. Doing so, the tensor-to-scalar ratio in our model can be
suppressed by the kinetic amplification factor in the bouncing phase as
described in the previous section.

\subsection{Loop quantum cosmology}

The realization of bouncing cosmologies becomes very natural in the frame
of loop quantum cosmology (LQC) since the classical big-bang singularity is
generically replaced by a quantum bounce when the space-time curvature of
the universe is of the order of the Planck scale \cite{Ashtekar:2006wn,
Singh:2009mz}. Several cosmological models have been studied in the context
of LQC, including inflation \cite{Ashtekar:2009mm, Linsefors:2012et,
Agullo:2013ai}, the matter bounce \cite{WilsonEwing:2012pu}, and the
ekpyrotic scenario \cite{Wilson-Ewing:2013bla}.  Note however that
anisotropies
are generically expected to become important near the bounce point ---with
the
exception of the ekpyrotic scenario--- and, while the bounce is robust in the
presence
of anisotropies \cite{Ashtekar:2009vc,Gupt:2012vi}, the analysis of the
cosmological perturbations becomes considerably more complex when
anisotropies
are important.

There are two realizations of the matter bounce scenario that have been
studied so far in LQC: the ``pure''  matter bounce model, where the dynamics
are matter-dominated at all times, including the bounce, and the new matter
bounce model (also called the matter-ekpyrotic bounce) where the space-time
is matter-dominated at the beginning of the contracting phase, while an
ekpyrotic scalar field dominates the dynamics during the end of the
contracting
era and also the bounce.

In LQC, the dynamics of cosmological perturbations are given by the effective
equation of motion for the Mukhanov-Sasaki variables that include quantum
gravity effects,
\begin{align} \label{lqc-eqs}
 {v^{i}_k}'' + \left( c_s^2 k^2 - \frac{z_i''}{z_i} \right) v^{i}_k = 0~,
\end{align}
where $k$ labels the Fourier modes and the index ${i=\{S, T \}}$ denotes the
scalar and tensor modes respectively. The detailed forms of the sound speed
parameter $c_s$ and the coefficient $z_i$ for holonomy-corrected LQC are
$c_s^2 = 1-2\rho/\rho_c$, $z_T^2=a^2/c_s^2$, and $z_S^2 = a^2(\rho+P)/H^2$.
Here $\rho_c \sim M_p^4$ is the critical energy density of LQC where the
bounce occurs.  These are the equations of motion that were used to determine
the observational predictions of the pure matter bounce and the
matter-ekpyrotic
bounce models in LQC.

In the pure matter bounce model in LQC, tensor perturbations are strongly
suppressed during the bounce due to the quantum gravity modification of
$z_T$ and this gives a predicted tensor-to-scalar ratio of
$r \sim \mathcal{O}(10^{-3})$, well below the signal detected by BICEP2
\cite{WilsonEwing:2012pu}.  In addition, the amplitude of the spectrum of
scalar perturbations is of the order of $\rho_c/M_p^4$, and therefore in
order to match observations, it is necessary for $\rho_c$ to be several
orders of magnitude below the Planck energy density.  This is problematic
as heuristic arguments relating LQC and the full theory of loop quantum
gravity indicate that $\rho_c$ is expected to be at most one or two orders
of magnitude below $\rho_{\rm Pl}$.

This last problem is avoided in the new matter bounce model for the following
reason: when the universe evolves into the ekpyrotic phase, all the
perturbation
modes at super-Hubble scales freeze and thus the amplitude of the
perturbations
are entirely determined by the value of the Hubble parameter at the beginning
of the ekpyrotic phase, $H_\mathrm{E}$ \cite{Cai:2014zga}. Because of this, the
observed
amplitude of the scalar perturbations determines $H_\mathrm{E}$, not $\rho_c$.
In addition, the ekpyrotic phase also dilutes the anisotropies before the
bounce occurs and hence the BKL instability is avoided in this model.
In order to determine how the recent results of the BICEP2 collaboration
constrain the matter-ekpyrotic bounce in LQC, it is necessary to determine
the amplitude of the primordial tensor fluctuations in this scenario.

The dynamics of scalar perturbations in the LQC matter-ekpyrotic bounce model
have been studied in detail in \cite{Cai:2014zga}, and this analysis is easy
to extend to tensor perturbations as their evolution is given by a very
similar differential equation as seen in Eq.\ \eqref{lqc-eqs}.  Due to the
fact that
their equations of motion are very similar at times well before the bounce,
the
amplitude of the spectra of the scalar and tensor modes are of the same
order,
and it is easy to check that if the ekpyrotic scalar field dominates the
dynamics
during the bounce,both the scalar and tensor modes evolve trivially through
the
bounce (note that this is very different to what happens if the matter field
dominates the dynamics during the bounce). The result is that, as in other
matter-ekpyrotic bounce models without entropy perturbations, the resulting
amplitude of the tensor perturbations is significantly larger than for the
scalar
perturbations and therefore this particular model is ruled out by
observations.

However, if there is more than one matter field then entropy perturbations
may
become important, and they have been neglected in the above analysis. As
explained
in Sec.\ \ref{s.matt-bounce}, entropy perturbations can significantly
increase the
amplitude of scalar perturbations, while not affecting the dynamics of tensor
perturbations in any way, thus decreasing the tensor-to-scalar ratio.
Therefore,
for the matter-ekpyrotic bounce model to be viable in LQC, it will be
necessary
to include entropy perturbations in some manner, perhaps as is done in the
new
matter bounce model presented in Sec.\ \ref{ss.new-mb}.

Finally, it is possible (at least for the flat FLRW space-time) to express
LQC as a teleparallel theory, which leads to slightly different equations of
motion for cosmological perturbations \cite{Haro:2013bea}. In this
setting, as there exist solutions with a large range of tensor-to-scalar
ratios, $r \in [0.1243, 13.4375]$ \cite{deHaro:2014kxa}, it is possible to
obtain a value of $r$ that is compatible with the results of the BICEP2
collaboration.

\section{Conclusion}
\label{s.concl}

In this work, we confronted various bouncing cosmologies with the recently
released BICEP2 data. In particular, we analyzed two scenarios in the
effective field theory framework, namely  the matter bounce curvaton scenario
and new matter bounce cosmology, and three modified gravity theories, namely
Ho\v{r}ava-Lifshitz gravity, the $f(T)$ theories, and loop quantum cosmology.
In all of these models, we showed their capability of generating primordial
gravitational waves.

Since matter bounce models typically produce a large amount of primordial
tensor fluctuations, specific mechanisms for their suppression are needed. In
the matter bounce curvaton scenario, introducing an extra scalar coupled to
the bouncing field induces a controllable amplification of the entropy modes
during the bouncing phase, and since these modes will be transferred into
curvature perturbations the resulting tensor-to-scalar ratio is suppressed to
a value in agreement with the observations of the BICEP2 collaboration.

Another possibility, called the new matter bounce cosmology, is to have two
scalar fields, one driving the matter contracting phase and the other driving
the ekpyrotic contraction and the nonsingular bounce. Thus, the entropy modes
are converted into curvature perturbations when the universe enters the
ekpyrotic phase before the bounce, and the resulting tensor-to-scalar ratio
is again suppressed to observed values.

Furthermore, in both of these models we used the BICEP2 and the Planck
results in order to constrain the free parameters in these models, namely the
energy scale of the bounce, the slope of the Hubble rate during the bouncing
phase, or the Hubble rate at the beginning of the ekpyrotic-dominated phase
for the new matter bounce cosmology.

Finally, we considered bouncing cosmologies in the framework of modified
gravity. In particular, in both the Ho\v{r}ava-Lifshitz bounce model and
the $f(T)$ gravity bounce, we have argued that the presence of a curvaton
field may suppress the tensor-to-scalar ratio to its observed values. We
leave the detailed analysis of this topic for a follow-up study.

In loop quantum cosmology, two realizations of the matter bounce have been
studied.  In the simplest matter bounce model where there is only one matter
field, the amplitude of the tensor perturbations is significantly diminished
during the bounce due to quantum gravity effects; this process predicts a
very small value of $r \sim \mathcal{O}(10^{-3})$, well below the value
observed by BICEP2.  The other model that has been studied is the new matter
bounce scenario, which in the absence of entropy perturbations predicts a
large amplitude for the tensor perturbations (in this case quantum gravity
effects do not modify the spectrum during the bounce).  Therefore, for the
new matter bounce scenario in LQC to be viable, it is also necessary to
include entropy perturbations in order to lower the value of $r$ to a value
in agreement with the results of BICEP2.  Also, as can be seen here, the
dominant field during the bounce significantly affects how the value of $r$
changes during the bounce and therefore it seems 
likely that by carefully choosing this field, it may be possible to obtain a
tensor-to-scalar ratio in agreement with observations.  We leave this
possibility for future work.

In summary, the predictions of the matter bounce cosmologies where entropy
perturbations significantly increase the amplitude of scalar perturbations
remain consistent with observations, and thus these models are good
alternatives to inflation.

\acknowledgments
We are indebted to Robert Brandenberger and Jean-Luc Lehners for valuable
comments. We also thank Taotao Qiu for helpful discussions in the initial
stages of the project. The work of YFC and JQ is supported in part by NSERC
and by funds from the Canada Research Chair program. The research of ENS is
implemented within the framework of the Action ``Supporting Postdoctoral
Researchers'' of the Operational Program ``Education and Lifelong Learning''
(Actions Beneficiary: General Secretariat for Research and Technology), and
is co-financed by the European Social Fund (ESF) and the Greek State.  The
work of EWE is supported in part by a grant from the John Templeton
Foundation. The opinions expressed in this publication are those of the
authors and do not necessarily reflect the views of the John Templeton
Foundation.


\begin{thebibliography}{99}
\raggedright

\bibitem{Ade:2014xna} 
  P.~A.~R.~Ade {\it et al.}  [BICEP2 Collaboration],
  Phys.\ Rev.\ Lett.\  {\bf 112}, 241101 (2014)
  [arXiv:1403.3985 [astro-ph.CO]].



\bibitem{Lizarraga:2014eaa} 
  J.~Lizarraga, J.~Urrestilla, D.~Daverio, M.~Hindmarsh, M.~Kunz and A.~R.~Liddle,
  Phys.\ Rev.\ Lett.\  {\bf 112}, 171301 (2014)
  [arXiv:1403.4924 [astro-ph.CO]].



\bibitem{Moss:2014cra} 
  A.~Moss and L.~Pogosian,
  Phys.\ Rev.\ Lett.\  {\bf 112}, 171302 (2014)
  [arXiv:1403.6105 [astro-ph.CO]].



\bibitem{Bonvin:2014xia} 
  C.~Bonvin, R.~Durrer and R.~Maartens,
  Phys.\ Rev.\ Lett.\  {\bf 112}, 191303 (2014)
  [arXiv:1403.6768 [astro-ph.CO]].



\bibitem{Grishchuk:1974ny} 
  L.~P.~Grishchuk,
  Sov.\ Phys.\ JETP {\bf 40}, 409 (1975)
  [Zh.\ Eksp.\ Teor.\ Fiz.\  {\bf 67}, 825 (1974)].



\bibitem{Starobinsky:1979ty} 
  A.~A.~Starobinsky,
  JETP Lett.\  {\bf 30}, 682 (1979)
  [Pisma Zh.\ Eksp.\ Teor.\ Fiz.\  {\bf 30}, 719 (1979)].



\bibitem{Rubakov:1982df} 
  V.~A.~Rubakov, M.~V.~Sazhin and A.~V.~Veryaskin,
  Phys.\ Lett.\ B {\bf 115}, 189 (1982).



\bibitem{Starobinsky:1985ww} 
  A.~A.~Starobinsky,
  Sov.\ Astron.\ Lett.\  {\bf 11}, 133 (1985).



\bibitem{Brandenberger:1988aj} 
  R.~H.~Brandenberger and C.~Vafa,
  Nucl.\ Phys.\ B {\bf 316}, 391 (1989).



\bibitem{Brandenberger:2006xi} 
  R.~H.~Brandenberger, A.~Nayeri, S.~P.~Patil and C.~Vafa,
  Phys.\ Rev.\ Lett.\  {\bf 98}, 231302 (2007)
  [hep-th/0604126].



\bibitem{Brandenberger:2014faa} 
  R.~H.~Brandenberger, A.~Nayeri and S.~P.~Patil,
  arXiv:1403.4927 [astro-ph.CO].



\bibitem{Biswas:2014kva} 
  T.~Biswas, T.~Koivisto and A.~Mazumdar,
  arXiv:1403.7163 [hep-th].



\bibitem{Wands:1998yp} 
  D.~Wands,
  Phys.\ Rev.\ D {\bf 60}, 023507 (1999)
  [gr-qc/9809062].



\bibitem{Finelli:2001sr} 
  F.~Finelli and R.~Brandenberger,
  Phys.\ Rev.\ D {\bf 65}, 103522 (2002)
  [hep-th/0112249].



\bibitem{Cai:2008qw} 
  Y.~F.~Cai, T.~t.~Qiu, R.~Brandenberger and X.~m.~Zhang,
  Phys.\ Rev.\ D {\bf 80}, 023511 (2009)
  [arXiv:0810.4677 [hep-th]].



\bibitem{Kehagias:2014wza} 
  A.~Kehagias and A.~Riotto,
  Phys.\ Rev.\ D {\bf 89}, 101301 (2014)
  [arXiv:1403.4811 [astro-ph.CO]].



\bibitem{Ma:2014vua} 
  Y.~Z.~Ma and Y.~Wang,
  arXiv:1403.4585 [astro-ph.CO].



\bibitem{Harigaya:2014qza} 
  K.~Harigaya and T.~T.~Yanagida,
  Phys.\ Lett.\ B {\bf 734}, 13 (2014)
  [arXiv:1403.4729 [hep-ph]].



\bibitem{Gong:2014qga} 
  J.~O.~Gong,
  JCAP {\bf 1407}, 022 (2014)
  [arXiv:1403.5163 [astro-ph.CO]].



\bibitem{Miranda:2014wga} 
  V.~Miranda, W.~Hu and P.~Adshead,
  Phys.\ Rev.\ D {\bf 89}, 101302 (2014)
  [arXiv:1403.5231 [astro-ph.CO]].



\bibitem{Hertzberg:2014aha} 
  M.~P.~Hertzberg,
  arXiv:1403.5253 [hep-th].



\bibitem{Lyth:2014yya} 
  D.~H.~Lyth,
  arXiv:1403.7323 [hep-ph].



\bibitem{Xia:2014tda} 
  J.~Q.~Xia, Y.~F.~Cai, H.~Li and X.~Zhang,
  Phys.\ Rev.\ Lett.\  {\bf 112}, 251301 (2014)
  [arXiv:1403.7623 [astro-ph.CO]].



\bibitem{Hazra:2014jka} 
  D.~K.~Hazra, A.~Shafieloo, G.~F.~Smoot and A.~A.~Starobinsky,
  arXiv:1404.0360 [astro-ph.CO].



\bibitem{Cai:2014bda} 
  Y.~F.~Cai, J.~O.~Gong and S.~Pi,
  arXiv:1404.2560 [hep-th].



\bibitem{Hossain:2014coa} 
  M.~W.~Hossain, R.~Myrzakulov, M.~Sami and E.~N.~Saridakis,
  Phys.\ Rev.\ D {\bf 89}, 123513 (2014)
  [arXiv:1404.1445 [gr-qc]].



\bibitem{Hu:2014aua} 
  B.~Hu, J.~W.~Hu, Z.~K.~Guo and R.~G.~Cai,
  Phys. Rev. D, volume 90, eid 023544, 2014
  [arXiv:1404.3690 [astro-ph.CO]].



\bibitem{Zhao:2014rna} 
  W.~Zhao, C.~Cheng and Q.~G.~Huang,
  arXiv:1403.3919 [astro-ph.CO].



\bibitem{Zhang:2014dxk} 
  J.~F.~Zhang, Y.~H.~Li and X.~Zhang,
  arXiv:1403.7028 [astro-ph.CO].



\bibitem{DiBari:2014oja} 
  P.~Di Bari, S.~F.~King, C.~Luhn, A.~Merle and A.~Schmidt-May,
  arXiv:1404.0009 [hep-ph].



\bibitem{Li:2014cka} 
  H.~Li, J.~Q.~Xia and X.~Zhang,
  arXiv:1404.0238 [astro-ph.CO].



\bibitem{Chung:2014woa} 
  Y.~C.~Chung and C.~Lin,
  JCAP {\bf 1407}, 020 (2014)
  [arXiv:1404.1680 [astro-ph.CO]].



\bibitem{Cai:2014hja} 
  Y.~F.~Cai and Y.~Wang,
  Phys.\ Lett.\ B {\bf 735}, 108 (2014)
  [arXiv:1404.6672 [astro-ph.CO]].



\bibitem{Hossain:2014ova} 
  M.~W.~Hossain, R.~Myrzakulov, M.~Sami and E.~N.~Saridakis,
  arXiv:1405.7491 [gr-qc].



\bibitem{Cai:2011zx} 
  Y.~F.~Cai, R.~Brandenberger and X.~Zhang,
  JCAP {\bf 1103}, 003 (2011)
  [arXiv:1101.0822 [hep-th]].



\bibitem{Cai:2012va} 
  Y.~F.~Cai, D.~A.~Easson and R.~Brandenberger,
  JCAP {\bf 1208}, 020 (2012)
  [arXiv:1206.2382 [hep-th]].



\bibitem{Cai:2013kja} 
  Y.~F.~Cai, E.~McDonough, F.~Duplessis and R.~H.~Brandenberger,
  JCAP {\bf 1310}, 024 (2013)
  [arXiv:1305.5259 [hep-th]].



\bibitem{Cai:2014bea} 
  Y.~F.~Cai,
  Sci.\ China Phys.\ Mech.\ Astron.\  {\bf 57}, 1414 (2014)
  [arXiv:1405.1369 [hep-th]].



\bibitem{Bessada:2012kw} 
  D.~Bessada, N.~Pinto-Neto, B.~B.~Siffert and O.~D.~Miranda,
  JCAP {\bf 1211}, 054 (2012)
  [arXiv:1207.5863 [gr-qc]].



\bibitem{Brandenberger:2009yt} 
  R.~Brandenberger,
  Phys.\ Rev.\ D {\bf 80}, 043516 (2009)
  [arXiv:0904.2835 [hep-th]].



\bibitem{Cai:2011tc} 
  Y.~F.~Cai, S.~H.~Chen, J.~B.~Dent, S.~Dutta and E.~N.~Saridakis,
  Class.\ Quant.\ Grav.\  {\bf 28}, 215011 (2011)
  [arXiv:1104.4349 [astro-ph.CO]].



\bibitem{Cai:2014zga} 
  Y.~F.~Cai and E.~Wilson-Ewing,
  JCAP {\bf 1403}, 026 (2014)
  [arXiv:1402.3009 [gr-qc]].



\bibitem{WilsonEwing:2012pu} 
  E.~Wilson-Ewing,
  JCAP {\bf 1303}, 026 (2013)
  [arXiv:1211.6269 [gr-qc]].



\bibitem{Cai:2011ci} 
  Y.~F.~Cai, R.~Brandenberger and X.~Zhang,
  Phys.\ Lett.\ B {\bf 703}, 25 (2011)
  [arXiv:1105.4286 [hep-th]].



\bibitem{Ade:2013zuv} 
  P.~A.~R.~Ade {\it et al.}  [Planck Collaboration],
  arXiv:1303.5076 [astro-ph.CO].



\bibitem{Cai:2013vm} 
  Y.~F.~Cai, R.~Brandenberger and P.~Peter,
  Class.\ Quant.\ Grav.\  {\bf 30}, 075019 (2013)
  [arXiv:1301.4703 [gr-qc]].



\bibitem{Osipov:2013ssa} 
  M.~Osipov and V.~Rubakov,
  JCAP {\bf 1311}, 031 (2013)
  [arXiv:1303.1221 [hep-th]].



\bibitem{Koehn:2013upa} 
  M.~Koehn, J.~L.~Lehners and B.~A.~Ovrut,
  Phys.\ Rev.\ D {\bf 90}, 025005 (2014)
  [arXiv:1310.7577 [hep-th]].



\bibitem{Cai:2009in} 
  Y.~F.~Cai and E.~N.~Saridakis,
  JCAP {\bf 0910}, 020 (2009)
  [arXiv:0906.1789 [hep-th]].



\bibitem{Calcagni:2009ar} 
  G.~Calcagni,
  JHEP {\bf 0909}, 112 (2009)
  [arXiv:0904.0829 [hep-th]].



\bibitem{Kiritsis:2009sh} 
  E.~Kiritsis and G.~Kofinas,
  Nucl.\ Phys.\ B {\bf 821}, 467 (2009)
  [arXiv:0904.1334 [hep-th]].



\bibitem{Cai:2009hc} 
  Y.~F.~Cai and X.~Zhang,
  Phys.\ Rev.\ D {\bf 80}, 043520 (2009)
  [arXiv:0906.3341 [astro-ph.CO]].



\bibitem{Linder:2010py} 
  E.~V.~Linder,
  Phys.\ Rev.\ D {\bf 81}, 127301 (2010)
  [Erratum-ibid.\ D {\bf 82}, 109902 (2010)]
  [arXiv:1005.3039 [astro-ph.CO]].



\bibitem{Chen:2010va} 
  S.~H.~Chen, J.~B.~Dent, S.~Dutta and E.~N.~Saridakis,
  Phys.\ Rev.\ D {\bf 83}, 023508 (2011)
  [arXiv:1008.1250 [astro-ph.CO]].



\bibitem{Dent:2011zz} 
  J.~B.~Dent, S.~Dutta and E.~N.~Saridakis,
  JCAP {\bf 1101}, 009 (2011)
  [arXiv:1010.2215 [astro-ph.CO]].



\bibitem{Ashtekar:2006wn} 
  A.~Ashtekar, T.~Pawlowski and P.~Singh,
  Phys.\ Rev.\ D {\bf 74}, 084003 (2006)
  [gr-qc/0607039].



\bibitem{Singh:2009mz} 
  P.~Singh,
  Class.\ Quant.\ Grav.\  {\bf 26}, 125005 (2009)
  [arXiv:0901.2750 [gr-qc]].



\bibitem{Ashtekar:2009mm} 
  A.~Ashtekar and D.~Sloan,
  Phys.\ Lett.\ B {\bf 694}, 108 (2010)
  [arXiv:0912.4093 [gr-qc]].



\bibitem{Linsefors:2012et} 
  L.~Linsefors, T.~Cailleteau, A.~Barrau and J.~Grain,
  Phys.\ Rev.\ D {\bf 87}, no. 10, 107503 (2013)
  [arXiv:1212.2852 [gr-qc]].



\bibitem{Agullo:2013ai} 
  I.~Agull\'o, A.~Ashtekar and W.~Nelson,
  Class.\ Quant.\ Grav.\  {\bf 30}, 085014 (2013)
  [arXiv:1302.0254 [gr-qc]].



\bibitem{Wilson-Ewing:2013bla} 
  E.~Wilson-Ewing,
  JCAP {\bf 1308}, 015 (2013)
  [arXiv:1306.6582 [gr-qc]].



\bibitem{Ashtekar:2009vc} 
  A.~Ashtekar and E.~Wilson-Ewing,
  Phys.\ Rev.\ D {\bf 79}, 083535 (2009)
  [arXiv:0903.3397 [gr-qc]].



\bibitem{Gupt:2012vi} 
  B.~Gupt and P.~Singh,
  Phys.\ Rev.\ D {\bf 86}, 024034 (2012)
  [arXiv:1205.6763 [gr-qc]].



\bibitem{Haro:2013bea} 
  J.~Haro,
  JCAP {\bf 1311}, 068 (2013)
  [Erratum-ibid.\  {\bf 1405}, E01 (2014)]
  [arXiv:1309.0352 [gr-qc]].



\bibitem{deHaro:2014kxa} 
  J.~de Haro and J.~Amor\'{o}s,
  arXiv:1403.6396 [gr-qc].

\end{thebibliography}
\end{document}